\documentclass[a4paper,11pt]{article}
\usepackage{pos}

\usepackage{braket}
\usepackage{tikz}
\usetikzlibrary{arrows.meta}
\usetikzlibrary{positioning}
\usetikzlibrary{decorations}
\usetikzlibrary{decorations.markings}

\DeclareMathOperator{\SU}{\mathrm{SU}}
\DeclareMathOperator{\Z}{\mathbb{Z}}

\newcommand{\intr}{\mathrm{intr}}

\title{Intrinsic width of the flux tube in 2+1 dimensional Yang-Mills theories}
\ShortTitle{Intrinsic width of the flux tube in 2+1D YM}

\author*[a]{Lorenzo Verzichelli}
\author[a]{Michele Caselle}
\author[a]{Elia Cellini}
\author[a]{Alessandro Nada}
\author[a]{Dario Panfalone}

\affiliation[a]{Universit\`a degli Studi di Torino, Via Pietro Giuria 1, Torino, Italy}

\emailAdd{lorenzo.verzichelli@unito.it}
\emailAdd{michele.caselle@unito.it}
\emailAdd{elia.cellini@unito.it}
\emailAdd{alessandro.nada@unito.it}
\emailAdd{dario.panfalone@unito.it}

\abstract{We study the shape of the flux tube in lattice Yang-Mills theories and in particular its intrinsic width. 
In the framework of the Effective String Theory description of the confining flux tube this intrinsic width has no measurable effects on the inter-quark static potential, but it can be precisely detected looking at the profile of the flux tube. We address this problem with a set of high precision simulations  in the (2+1) dimensional $\SU(2)$ model. We find two different behaviours as a function of the temperature. In the low temperature regime ($T \ll T_c$) we find a good agreement with an expression inspired by the dual superconductive model of confinement. In the high temperature regime ($T \lesssim T_c$) our data agree with  a model based on the Svetitsky-Yaffe mapping. All our data in this regime can be described in terms of only one length scale, the intrinsic width, which turns out to be the same scale appearing in the confining inter-quark static potential.}

\FullConference{%
  The 41st International Symposium on Lattice Field Theory (LATTICE2024)\\
  28 July - 3 August 2024\\
   Liverpool, UK
}

\begin{document}
\maketitle

\section{Introduction}
A remarkable feature of confining gauge theories is the formation of a flux tube between two charged (or colored) sources. 
In the Effective String Theory (EST) description of confinement, this flux tube is described as a vibrating string connecting the quark and the anti-quark~\cite{Luscher:1980ac, Luscher:1980fr}. 
This string is assumed to be width-less but acquires a finite thickness due to quantum fluctuations~\cite{Luscher:1980iy}. This finite width can be studied precisely in the EST framework and shows a rather non trivial behaviour. If one assumes a Nambu-Goto action for the EST then it can be shown
that at low temperatures the square width must increase logarithmically with the inter-quark distance~\cite{Luscher:1980iy}. This prediction has been confirmed by numerical simulations in various pure gauge theories (see for 
instance~\cite{Caselle:1995fh, Zach:1997yz, Koma:2003gi,Gliozzi:2010zv,Bakry:2010zt}). At high temperature, but still in the confining phase, the dependence of the square width on the inter-quark distance becomes linear with a proportionality constant which diverges as the deconfinement transition is approached \cite{Allais:2009uos,Caselle:2010zs}.  Also these predictions were
nicely confirmed by numerical simulations \cite{Caselle:2010zs,Gliozzi:2010jh}.

However we know that the Nambu-Goto action is not the correct EST description of the inter-quark potential.
Thanks to the remarkable universality theorems discussed in~\cite{Luscher:2004ib,Dubovsky:2012sh,Aharony:2013ipa}
 we know that the first few perturbative orders in the large distance expansion of the effective string action are universal and coincide with
those of the Nambu-Goto action (and this explains the good agreement between predictions and simulations), but corrections appear at higher order. 

These deviations from the Nambu-Goto action may be of two types: they can be due to higher order terms in the large distance expansion of the effective string action (for a detailed discussion see for instance~\cite{Caselle:2024zoh}) or they can be due to the coupling of the massless degrees of freedom of the effective string with massive (non-stringy) "intrinsic" excitations of the theory (for a discussion in the case of the three dimensional U(1) LGT see \cite{Aharony:2024ctf}). Due to our incomplete understanding of how the effective string emerges from QCD we have no precise description of these massive modes, of their dynamical origin and of their action, but we have a few hints which may help our intuition. The most important one is that these massive modes should manifest themselves as a sort of "\textit{intrinsic width}" of the flux tube, different from the width due to the quantum fluctuations discussed above (which we shall call in the following "\textit{effective width}" to avoid confusion). In this respect the intrinsic width, which we shall denote in the following as $\lambda$, can be viewed as the residual thickness of the flux tube when the inter-quark distance $d$ is pushed down to the scale (typically $d \sim 1/\sqrt{\sigma})$ below which the effective string description does not hold any more and does not contribute to the flux tube thickness.

From a numerical point of view the evaluation of this intrinsic width is a rather non trivial task since in the above formulation of the effective width (both in the low and in the high temperature regimes) it is hidden in the additive constants contained in the corresponding expressions. 
The most effective way to measure it is to look at the \textit{shape} of the flux tube (see for instance \cite{Cardoso:2012aj, Cardoso:2013lla, Baker:2018mhw, Baker:2023dnn}).  This shape is predicted to be Gau\ss ian in the Nambu-Goto case (a prediction which was recently confirmed numerically in ref.~\cite{Caselle:2024ent}) and we can find hints of the intrinsic width looking at the deviation from this Gau\ss ian shape of the actual flux tube in LGTs. 

In this contribution we report some preliminary result in this direction. We studied in particular the $\SU(2)$ model in (2+1) dimensions which represents a perfect laboratory to address this issue, since it shares the same infrared  behaviour of more complex four dimensional LGTs but is much simpler to simulate and allows to reach precise determinations of the flux tube shape with a contained numerical effort.

\section{Profile of the flux tube on the lattice}
We studied the $\SU(2)$ Yang-Mills theory in 2+1 space-time dimensions with the standard Wilson action. We simulated the model for a few different values of $\beta$ to test the scaling behaviour of our results, on cubic lattices of size $N_s$ in the spatial directions and $N_t$ in the (compactified) Euclidean time direction. For each $\beta$ we simulated the model at different temperatures, always in the confined phase, $T < T_c$, where $T_c$ is the deconfinement temperature. In the low temperature regime, for $T < 0.5 \, T_c$, we used the L\"uscher-Weisz multilevel method.
In these cases we updated the configurations $O(10^2)$ times in the outermost level and $O(10^4)$ in the innermost one. In the high temperature regime,
for $T > 0.5 \, T_c$ the configurations were updated with the standard algorithm, with a number of updates varying from $O(10^5)$ to $O(10^6)$, as the critical temperature was approached. We summarize in table \ref{tab1} some information on the simulations that we performed.

\begin{table}
\begin{center}

\begin{tabular}{|c|c|c|c|c|}

\hline
$\beta$     & $N_t$ & $N_s$ & $T / T_c$ & $d / a$ range \\
\hline
10.865412   & 96    & 96    & 0.12      & $9$           \\
\hline
 8.7683296  & 24    & 75    & 0.23      & $[ 9; 15]$    \\
10.865412   & 30    & 96    & 0.23      & $[ 9; 15]$    \\
\hline
10.865412   & 20    & 96    & 0.35      & $[ 9; 15]$    \\
\hline
10.865412   & 10    & 96    & 0.70      & $[11; 21]$    \\
11.9139532  & 11    & 96    & 0.70      & $[11; 19]$    \\
12.9624944  & 12    & 96    & 0.70      & $[11; 17]$    \\
14.011      & 13    & 120   & 0.70      & $[ 9; 21]$    \\
\hline
10.865412   &  8    & 96    & 0.87      & $[ 9; 21]$    \\
13.42445    & 10    & 120   & 0.87      & $[ 9; 21]$    \\
\hline

\end{tabular}
\end{center}
\caption{Some information on our simulations. The value of the critical temperatures as a function of $\beta$ was obtained using the scale setting reported in ref.~\cite{Liddle:2008kk}.}
\label{tab1}
\end{table}

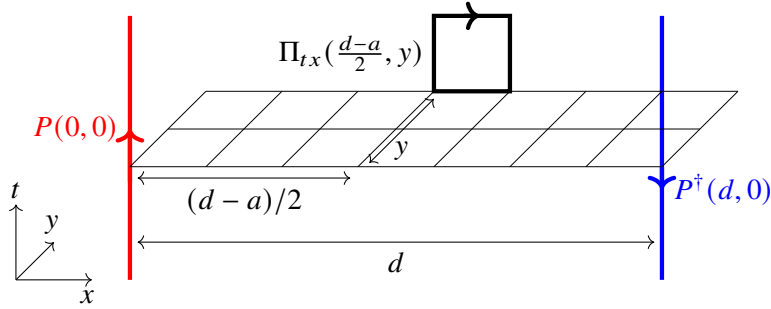
\begin{figure}[h]
    \centering
    \tikzset{
    	lattice point/.style={
    		draw,
    		fill,
    		circle,
    		minimum size=1.0mm,
    		inner sep=0,
    	},
    	intermediate point/.style={
    		draw,
    		fill,
    		circle,
    		minimum size=0.0mm,
    		white,
    		inner sep=0,
    	},    
    }
    
    \begin{center}
    	\begin{tikzpicture}[scale=1., transform shape]
    		
    		\draw [ -> ] (0, 0) -- (0, 1) node [pos=0.95, anchor=south] {$t$};
    		\draw [ -> ] (0, 0) -- (1, 0) node [pos=0.95, anchor=north] {$x$};
    		\draw [ -> ] (0, 0) -- (0.5, 0.5) node [pos=0.95, anchor=south] {$y$};
    		
    		\draw [red, ultra thick, ->] (1.5, 0) -- (1.5, 2.) node [anchor=east] {$P(0, 0)$};
    		\draw [red, ultra thick] (1.5, 1.3) -- (1.5, 3.5);
    		
    		\draw [blue, ultra thick] (8.5, 0) -- (8.5, 1.5);
    		\draw [blue, ultra thick, ->] (8.5, 3.5) -- (8.5, 1.2) node [anchor=west] {$P^\dagger(d, 0)$};
    		
    		\draw [ultra thin] (1.5, 1.5) -- (8.5, 1.5);
    		\draw [ultra thin] (2.0, 2.0) -- (9.0, 2.0);
    		\draw [ultra thin] (2.5, 2.5) -- (9.5, 2.5);
    		
    		\draw [ultra thin] (1.5, 1.5) -- (2.5, 2.5);
    		\draw [ultra thin] (2.5, 1.5) -- (3.5, 2.5);
    		\draw [ultra thin] (3.5, 1.5) -- (4.5, 2.5);
    		\draw [ultra thin] (4.5, 1.5) -- (5.5, 2.5);
    		\draw [ultra thin] (5.5, 1.5) -- (6.5, 2.5);
    		\draw [ultra thin] (6.5, 1.5) -- (7.5, 2.5);
    		\draw [ultra thin] (7.5, 1.5) -- (8.5, 2.5);
    		\draw [ultra thin] (8.5, 1.5) -- (9.5, 2.5);
    		
    		\draw [ultra thick] (5.5, 2.5) -- (6.5, 2.5) -- (6.5, 3.5) -- (5.5, 3.5) -- cycle;
    		\draw [ultra thick, ->] (5.9, 3.5) -- (6.1, 3.5);
    		\node at (5.5, 3) [anchor=east] {$\Pi_{tx}(\frac{d - a}{2}, y)$};
    		
    		\draw [ <-> ] (1.6, 0.5) -- (8.4, 0.5) node [pos=0.5, anchor=north] {$d$};
    		\draw [ <-> ] (1.6, 1.35) -- (4.4, 1.35) node [pos=0.5, anchor=north] {$(d - a) / 2$};
    		\draw [ <-> ] (4.65, 1.55) -- (5.5, 2.4) node [pos=0.5, anchor=north] {$y$};
    		
    	\end{tikzpicture}
    \end{center}
    \caption{The three point function $F(d, y)$ of eq.~\eqref{eq:three_pts} discussed in the text. The thick lines denote the Polyakov loops.}
    \label{fig:poly_poly_plaq}
\end{figure}

In order to study the shape of the flux tube, we considered the correlation of two Polyakov loops oriented in opposite directions with a plaquette located in the plane orthogonal to the inter-quark axis, in the mean position between the two loops (see figure \ref{fig:poly_poly_plaq}).
We denote the distance between the two loops by $d$ and assume them to be separated along the $\hat{x}$ direction. The transverse displacement of the plaquette from the plane containing the two Polyakov loops is chosen to be the $y$ coordinate of the plaquette. The other coordinate in the plane the plaquette lies on is irrelevant due to translation invariance in the Euclidean time.
The plaquette can be chosen in different orientations. However it is known, see for example ref.~\cite{Bonati:2020orj}, that the most prominent signal is obtained considering the plaquette that corresponds to the chromo-electric field in the direction of the separation between the two Polyakov loops, \textit{i.e.} the plaquette in the $\hat{x}$-$\hat{t}$ plane. In order to have the plaquette exactly equidistant from the two Polyakov loops, in this setup, we choose their distance $d$ to be odd in units of the lattice spacing. The parallel (along the $\hat{x}$ direction) displacement of the plaquette from each Polyakov loop will thus be $(d - a) / 2$, where $a$ is the lattice spacing.

The three point function is thus defined as
\begin{equation}
    F(d, y) = \left< P(0, 0) \, \Pi_{tx} \! \left( \frac{d - a}{2}, y, t \right) \, P^\dagger \! (d, 0) \right>,
    \label{eq:three_pts}
\end{equation}
where $P(\Vec{x})$ is the trace of the Polyakov loop in the (2D) space point $\Vec{x}$ and $\Pi_{\mu\nu}(x)$ is the trace of the plaquette based on the (3D) space-time point $x$, extended along the positive $\hat{\mu}$ and $\hat{\nu}$ axes.

In the following we shall be interested in the connected component of the three points function, normalized by the correlation of the two Polyakov loops. We shall denote this connected component as $\rho(d,y)$ where, as mentioned above, $d$ is the inter-quark distance and $y$ the transverse displacement.
\begin{equation}
    \rho(d, y) = \frac{F(d, y)}{\braket{P(0, 0) \, P^\dagger(d, 0)}} - \braket{\Pi_{tx}}
\end{equation}

For each value of $d$, we will call \textit{profile} of the flux tube $\rho$ as a function fo $y$.

\section{Low temperature results}

\begin{figure}[h]
    \centering
    \includegraphics[width=0.75\linewidth]{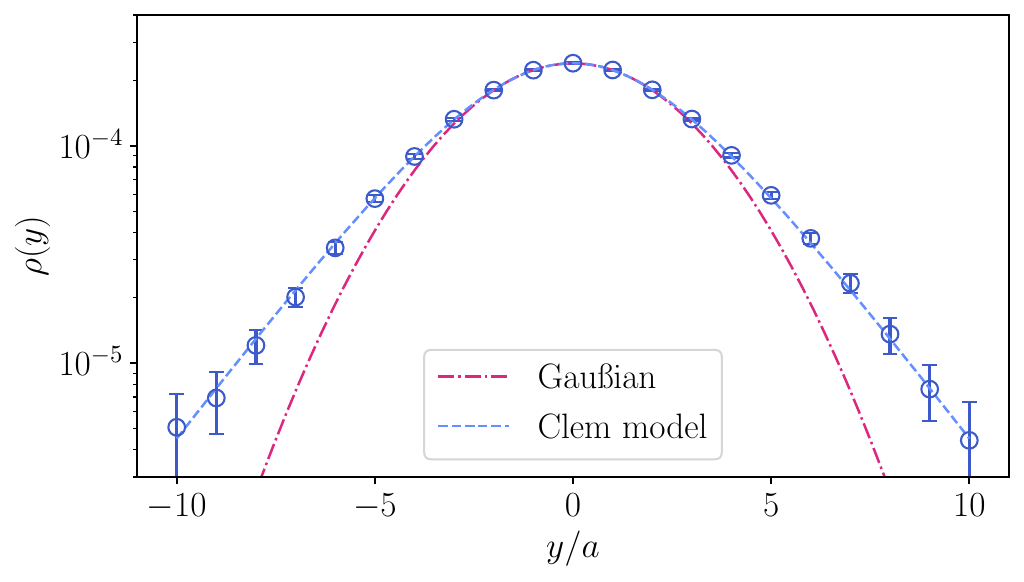}
    \caption{Profile at low temperature ($0.23 \, T_c$), obtained witha a simulation at $\beta = 10.865412$, on a $96 \times 96 \times 30$ lattice, The Gau\ss ian fit through the seven central points does not match the tails at large $y$. We also superimpose to our data a fit obtained with the model in \eqref{eq:clem_formula}.}
    \label{fig:non_gauss_lowT}
\end{figure}

In figure \ref{fig:non_gauss_lowT} we plot in a log scale the profile of the flux tube at $\beta=10.865412$ and $N_t=30$ (which corresponds to $T = 0.23 T_c$) and compare it with a Gau\ss ian profile (obtained in this particular case by fitting the data in the range $y/a \in [-3; 3]$ only). 

It is quite evident that the numerical data deviate from a Gau\ss ian distribution. If one is to extend a Gau\ss ian fit to a larger range of the transverse distance, for example $y/a \in [-10; 10]$, the $\chi^2$ becomes an order of magnitude larger than the degrees of freedom, making it clear that a Gau\ss ian model is not suitable to fit our data.

Looking at the figure we see that at large transverse displacement $y \gtrsim \sqrt{\sigma_0}$, the logarithm of the profile decays linearly with $y$. This implies that for large transverse displacement the profile decays approximately as a simple exponential $\exp(-|y| / \lambda)$. As discussed above this is the expected behaviour in presence of an intrinsic width  $\lambda$.

It would be interesting to find a meaningful way to interpolate between the Gau\ss ian behaviour at short distance and the exponentially decreasing behaviour at large distances. An interesting proposal in this respect is represented by the so called "Clem formula" which was introduced long ago by Clem  \cite{Clem1975} to describe Abrikosov vortices in superconductivity and was recently proposed (in the framework of the so called "dual superconductor" model of confinement) in the LGT context in \cite{Cea:2012qw} as a way to describe the flux tube shape. 

In the last few years this formula has been used to fit the profile of the flux tube in several works, both in non-Abelian gauge theories \cite{Cea:2012qw} and in full QCD (with dynamical fermions) \cite{Cea:2017ocq}. The Clem formula can be expressed as follows:
\begin{equation}
    \rho(y) = A \, K_0 \left( \frac{\sqrt{y^2 + {\xi_v}^2}}{\lambda} \right) \approx A \sqrt{\frac{\pi \, \lambda}{2 \, y}} \; \exp \left( -\frac{y}{\lambda} \right),
    \label{eq:clem_formula}
\end{equation}
where $K_0$ is the modified Bessel function of the second kind of order zero. The last term in eq.~\ref{eq:clem_formula} is obtained using the large distance expansion of the Bessel function, and makes it manifest the exponential decrease of the $\rho(y)$ for large values of $y$. On the contrary for small values of $y$, by expanding the square root we immediately find a Gau\ss ian behaviour as observed in the data. 
The Clem formula features two dimensionful parameters: a variational length $\xi_v$ and $\lambda$, which plays the role of the London length of the dual superconductor. $\lambda$ is the length scale that we interpret as the intrinsic width, while $\sqrt{\lambda \xi_v}$ is the width of the Gau\ss ian peak of the profile.

Our low temperature data show a good agreement with fits performed with the Clem formula. Examples of our results are shown in table \ref{tab:lowT_results}. We observe that the London length $\lambda$ does not depend on the distance between the two Polyakov loops, as it is generally expected for the intrinsic width of the flux tube.
As for the variational length $\xi_v$, it exhibits an evident growth with the distance $d$ between the loops. This behavior is consistent with expectations, since, as we discussed above, the flux tube is known to broaden as the inter-quark distance increases. However this behavior of the variational length lacks, to our knowledge, a clear explanation in the dual superconductor model.

As an alternative to the Clem formula we also tried two other fitting strategies. First we fitted the data with a simple convolution of a Gau\ss ian with an exponentially decreasing function\footnote{We thank M.~Pepe and O.~Aharony for this suggestion.}. Second, we tried to fit only the tails ($|y| \geq y_\intr$) (setting a threshold $y_\intr$ below which we neglected the data) of the profile with a pure exponential decrease. The values of the intrinsic width obtained with these strategies are also reported in table \ref{tab:lowT_results} as $\lambda^{\text{conv}}$ for the convolution fit and $\lambda^{\text{exp}}$ for the pure exponential fit (together with the value of the cut-off $y_\intr$).

\begin{table}[h]
    \centering
    \begin{tabular}{|c|cccc|c|cc|}
 \hline
 $d/a$ & $\lambda^{\text{clem}} / a$ & $A$ &      $\xi_v$ & $\chi^2 / n.dof$ & $\lambda^{\text{conv}} / a$ & $y_\intr$ & $\lambda^{\text{exp}} / a$ \\
 \hline
     9 &  1.893(31) & 0.001558(98) &    1.647(50) &    34.69/18 & 2.046(21) & 3       & 2.16(16) \\
    11 &  1.857(62) &  0.00314(55) &     2.33(15) &    16.47/18 & 2.053(67) & 4       & 1.98(17) \\
    13 &  1.89(12)  &   0.0048(18) &     2.77(32) &    18.23/18 & 2.083(72) & 4       & 2.17(20) \\
    15 &  1.86(19)  &   0.0102(79) &     3.51(69) &     9.12/18 & 2.12(14)  & 5       & 2.32(27) \\
    \hline
    \end{tabular}
    \caption{Results for $\lambda$ obtained with the three fit strategies, for  $\beta = 10.865412$ and $N_t=30$ (i.e. $T = 0.23 \, T_c$). $\lambda^{\text{conv}}$ corresponds to the fit with the convolution of a Gau\ss ian and an exponential; $\lambda^{\text{exp}}$ was obtained fitting with a pure exponential, only the data with $|y| \geq y_\intr$; $\lambda^{\text{clem}}$ is results from a fit with with the Clem formula. $\lambda^{\text{conv}}$ and $\lambda^{\text{exp}}$ are systematically larger than $\lambda^{\text{clem}}$. This is most probably due to the different power law prefactor in the two expressions. Notice that the value of $y_\intr$ increases with the inter-quark distance.}
    \label{tab:lowT_results}
\end{table}

We have no theoretical argument, nor sufficient numerical precision, to rule out any of these proposals, thus we consider the discrepancy between the values of $\lambda$ obtained with the different approaches, as the systematic error of our determination of the intrinsic width of the flux tube
in this regime.
Remarkably enough, the values of 
$\lambda$ that we obtain in this way show a good scaling behaviour and a negligible dependence on $T$  (at least up to $T = 0.35 \, T_c$, the discrepancies are less than $10 \%$, thus of the order of our error bars). Thus we consider our results as a reliable estimate of the intrinsic witdh of the confining flux tube in the low temperature regime ($T<T_c/2$). Converting our estimates in units of $\sqrt{\sigma_0}$ we end up with a final estimate of $\lambda\sqrt{\sigma_0}\sim 0.25$.



\section{High temperature results and the Svetitsky--Yaffe mapping}

At high temperature, fits performed with the Clem model lead to unsatisfactory values of the $\chi^2$. Furthermore, the fitted exponential decay of the tails becomes much more unstable and exhibits an apparent dependence on the distance between the two Polyakov loops. These observations suggest that the actual expression for the shape should be characterized in this regime by a stronger power low prefactor.
Such a power decay can, indeed, be predicted using the Svetitsky--Yaffe (SY) mapping \cite{Svetitsky:1982gs}, which relates correlation functions of $\SU(N)$ gauge theories to those of $\Z_N$-symmetric spin models in one fewer dimensions, near a second order phase transition. In particular, the $\SU(2)$ gauge theory in 2+1 dimensions presents a second order deconfining phase transition, in the proximity of which its correlation functions are related to those of the two dimensional Ising model, which can be studied analytically.
According to the SY mapping the three point function we considered is related to the spin-spin-energy correlator of the Ising model. This has been studied in ref.~\cite{Caselle:2006wr, Caselle:2012rp}, providing a model to fit the flux tube profile in the proximity of the deconfinement phase transition:
\begin{equation}
	\rho(y) = \frac{2 \pi \, A \, d}{K_0\{d / (2 \, \lambda)\}} \, \frac{1}{4 y^2 + d^2} \, \exp\left( -\frac{\sqrt{4 \, y^2 + d^2}}{2 \lambda}\right).
    \label{eq:SYmp_model}
\end{equation}
Remarkably, aside from the amplitude, the only dimensionful parameter present in the model is $\lambda$, which also in this case controls the exponential decay at large transverse displacement $y \gg d$. The role played by the variational length $\xi_v$ in the Clem formula is here played by $d$, which is the distance between the two Polyakov loops.
\begin{figure}
    \centering
    \includegraphics[width=0.75\linewidth]{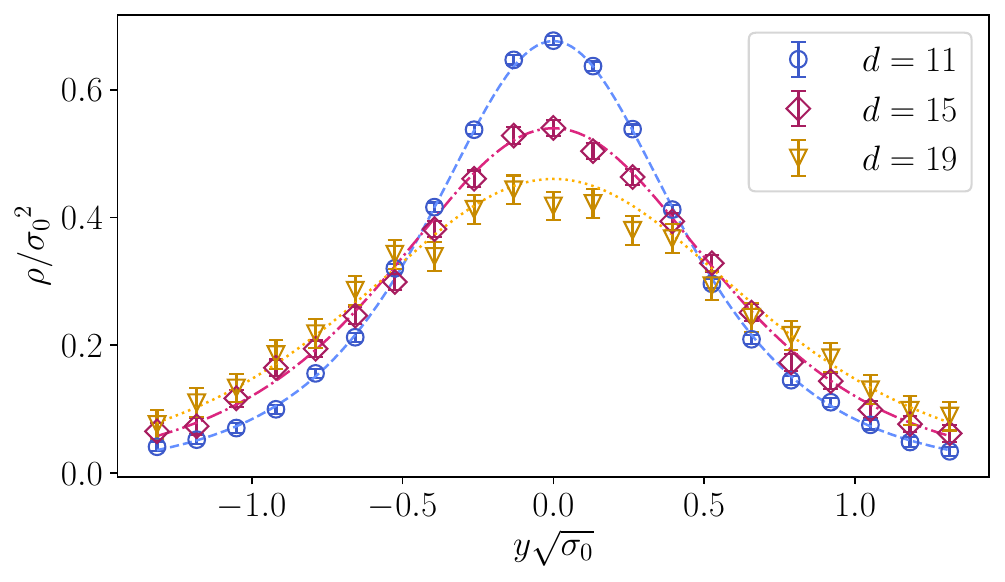}
    \caption{Fit to the profile at $T = 0.7 T_c$ using the model based on the SY mapping. The profile was obtained using an inverse coupling $\beta = 10.87$ on a $96 \times 96 \times 10$ lattice, for different values of $d$. The lines are all drawn assuming the same values of the free parameters of the model.}
    \label{fig:profile_highT_fit}
\end{figure}
The fits performed with the model in equation \eqref{eq:SYmp_model} are in perfect agreement with the data already for temperatures as low as $T = 0.7 \, T_c$, as shown by the values of the $\chi^2$ obtained fitting each profile separately. An example of these results is reported in table \ref{tab:highT_resilts}. We obtained also remarkably good $\chi^2$ values fitting all the data obtained for a given value of $\beta$ and $T$, for all the distances between the loops larger than a minimal threshold related to the critical radius of the EST (see the last two lines of table \ref{tab:highT_resilts}). Also in this case the values we obtain for $\lambda$ do not depend on the distance between the Polyakov loops.
Moreover, once it is expressed in units of the zero temperature string tension the intrinsic width has the expected scaling behaviour with the lattice spacing: $\sqrt{\sigma_0} \, \lambda$ does not depend on $\beta$ within statistical uncertainties.

\begin{table}
    \centering
\begin{tabular}{|ccccc|}
\hline
\multicolumn{5}{|l|}{$\beta = 10.865$, $a \sqrt{\sigma_0} = 0.13137(86)$, $N_t = 10$, $N_s = 96$} \\
\hline
 $d/a$ & $10^3 \, A$ & $\lambda / a$ & $\chi^2 / ndof$ &  $p_{val}$ \\
\hline
    11 & 1.747(29) &    4.62(16) &   105.13/94 & 20.33 \% \\
    13 & 1.732(39) &    4.62(21) &   107.53/94 & 16.08 \% \\
    15 & 1.714(51) &    4.60(28) &    83.96/94 & 76.15 \% \\
    17 & 1.709(68) &    4.71(39) &   101.53/94 & 27.97 \% \\
    19 & 1.800(96) &    5.43(62) &    84.81/94 & 74.03 \% \\
\hline
$>10$ & 1.747(27) &    4.65(14) &  571.12/574 &  52.61\% \\
$>12$ & 1.738(36) &    4.70(20) &  465.56/478 &  64.96\% \\
\hline
\multicolumn{2}{|c}{from $PP^\dagger$ correlator}&  4.5192(98) && \\
\hline
    \end{tabular}
    \caption{Fit results for the profile using the model based on the SY mapping. We also report the results of combined fits and the value of $\lambda$ that we extracted from the correlation of Polyakov loops.}
    \label{tab:highT_resilts}
\end{table}

A stringent cross-check of the model is that the parameter $\lambda$ is predicted in~\cite{Caselle:2006wr, Caselle:2012rp} to be exactly half the typical decay length of the Polyakov loop correlator: $2 \, \lambda = T / \sigma(T)$, where $\sigma(T)$ is the temperature-dependent string tension. Using the data for $\sigma(T)$ reported  in ref.~\cite{Caselle:2024zoh} we could check that our values of $\lambda$ perfectly agree with this prediction and, accordingly, that the intrinsic width $\lambda$ grows approaching the critical temperature (since the temperature-dependent string tension vanishes at the critical point). In particular we found, expressing the intrinsic width in units of the zero temperature string tension, at $T = 0.70 T_c$, $\lambda \, \sqrt{\sigma_0} = 0.597(10)$; while at $T = 0.87 T_c$, $\lambda \, \sqrt{\sigma_0} = 1.298(21)$, to be compared with the value $\lambda \, \sqrt{\sigma_0} \sim 0.25$ found in the low temperature regime.

\section{Conclusions}

We presented the results of numerical simulations for the profile of the flux tube, in the confining phase of the 2+1 dimensional $\SU(2)$ pure gauge theory, both at low and high temperature. In both regimes we identified  a new scale of the model, which controls the exponentially decreasing tails of the shape of flux tube. This length does not depend on the inter-quark distance and represents the so called  "intrinsic width" of the flux tube. 
In the low temperature regime the shape is well fitted by the Clem formula while at high temperature a model based on the SY mapping fits our data with one free parameter less with remarkable accuracy. In this model, the intrinsic width is related to the characteristic scale of the correlation between Polyakov loops, which is in turn the temperature divided by the temperature-dependent string tension. This relation is numerically verified within the statistical precision in the cases we examined. 

\section*{Acknowledgments}

We thank Ofer Aharony, Andrea Bulgarelli, Marco Panero and Michele Pepe for useful discussions.  The numerical simulations were run on the CINECA machines, according to the agreement between INFN and CINECA, under the projects of the SFT Scientific Initiative of INFN, whose support is acknowledged by all authors. The work was partially supported by the Simons Foundation through grant 994300 (Simons Collaboration on Confinement and QCD Strings). A.~Nada acknowledges support from the European Union - Next Generation EU, Mission 4 Component 1, CUPD53D23002970006, under the Italian PRIN “Progetti di Ricerca di Rilevante Interesse Nazionale – Bando 2022” prot. 2022ZTPK4E.

\bibliographystyle{JHEP}
\providecommand{\href}[2]{#2}\begingroup\raggedright\endgroup

\end{document}